\documentclass{emulateapj}

\newcommand{\mincir}{\raise
-2.truept\hbox{\rlap{\hbox{$\sim$}}\raise5.truept\hbox{$<$}\ }}
\newcommand{\magcir}{\raise
-2.truept\hbox{\rlap{\hbox{$\sim$}}\raise5.truept\hbox{$>$}\ }}
\newcommand{\minmag}{\raise
-2.truept\hbox{\rlap{\hbox{$<$}}\raise6.truept\hbox{$<$}\ }}

\shorttitle{The clustering and the environment of EROs}
\shortauthors{Georgakakis et al.}

\begin{document}

\title{The Phoenix Deep Survey: the clustering and the environment 
of Extremely Red Objects}

\author{
A. Georgakakis\altaffilmark{1}, 
J. Afonso\altaffilmark{2}, 
A. M. Hopkins\altaffilmark{3}, 
M. Sullivan\altaffilmark{4}, 
B. Mobasher\altaffilmark{5},
L. E. Cram\altaffilmark{6}
}

\altaffiltext{}{Based on observations collected at the European Southern
  Observatory, Chile, ESO 66.A-0193(A).}

\altaffiltext{1}{Institute of Astronomy \& Astrophysics,
  National Observatory of Athens, I. Metaxa \& B. Pavlou, Penteli,
  15236, Athens, Greece}
  \email{age@astro.noa.gr} 

\altaffiltext{2}{Centro de Astronomia da Universidade de Lisboa,
   Observat\'orio Astron\'omico de Lisboa, Tapada da Ajuda,
   1349-018 Lisboa, Portugal}

\altaffiltext{3}{Hubble Fellow, Department of Physics and Astronomy,
   University of Pittsburgh, 3941 O'Hara Street, Pittsburgh, PA 15260,
   USA} 

\altaffiltext{4}{Department of Astronomy and Astrophysics, University
of Toronto, 60 St. George Street, Toronto, ON M5S 3H8, Canada}  
  
\altaffiltext{5}{Space Telescope Science Institute, 3700 San Martin
Drive, Baltimore, MD 21218, USA}

\altaffiltext{6}{The Australian National University,  Canberra ACT
0200, Australia}




\begin{abstract}
In this paper we explore the clustering properties and the environment
of the Extremely Red Objects (EROs; $I-K>4$\,mag)  detected in a
$\approx180 \rm \, arcmin^2$ deep ($Ks\approx20$\,mag)
$Ks$-band survey of a region within the Phoenix Deep Survey, an
on-going  multiwavelength program aiming to investigate the nature and
the evolution of faint radio sources. Using our complete sample
of 289 EROs brighter than $Ks=20$\,mag we estimate a statistically 
significant ($\approx3.7\sigma$) angular correlation function 
signal with amplitude $A_w=8.7^{+2.1}_{-1.7}\times10^{-3}$ (assuming 
$w(\theta)=A_w\,\theta^{-0.8}$, with $\theta$ in deg), consistent with
earlier work based on smaller samples. This amplitude suggests a
clustering  length in the range $r_{o} = 12 - 17 \, h^{-1} \rm \,Mpc$,
implying that EROs trace regions of enhanced density. Using a novel
method we further explore the association of EROs with galaxy
overdensities by smoothing the $K$-band galaxy distribution using the
matched  filter algorithm of Postman et al. (1996) and then
cross-correlating the resulting density maps with the ERO
positions. Our analysis provides direct evidence  that EROs are
associated with overdensities at redshifts $z\ga1$.  We also exploit
the deep radio 1.4\,GHz data (limiting flux $60\mu$Jy) available to
explore the association of EROs and faint radio  sources and whether
the two populations trace similar large scale
structures. Cross-correlation of the two samples (after excluding  17
EROs with radio counterparts) gives a $2\sigma$ signal only  for the
sub-sample of high-$z$ radio sources ($z>0.6$). Although the
statistics are poor this  suggests that it is  the high-$z$ radio
sub-sample that traces similar structures with EROs.    
\end{abstract}


\section{Introduction}\label{sec_intro}
The study of the class of Extremely Red Objects (EROs; $I-K\ga4$\,mag;
e.g. Cimatti et al. 2002; Roche et al. 2002, 2003) has gained
significant impetus over the last few years with the realisation that
they can provide valuable information on galaxy evolution and
formation scenarios (c.f. Zepf 1997; Barger 1999; Rodighiero et
al. 2001; Daddi et al. 2000; Roche et al. 2002; V\"{a}is\"{a}nen \&
Johansson 2004a, b). The very red colours of these objects suggest
that they comprise evolved galaxies at redshifts $z\ga1$, the
progenitors of present-day ellipticals. Study of the properties of
such high-$z$ systems can indeed provide tight constraints on
competing elliptical galaxy formation scenarios: monolithic collapse
early in the Universe ($z_f>2-3$) followed by passive evolution
(e.g. Eggen et al. 1962; Larson 1975) versus hierarchical merging and
relatively recent formation epochs (Baugh et al. 1996; Kauffmann
1996).

Spectroscopic follow-up observations of either complete ERO samples
(Cimatti et al. 2002; Cimatti et al. 2003) or individual  systems
(Dunlop et al. 1996; Spinrad et al. 1997; Stanford et al. 1997) have
indeed confirmed that a large fraction of EROs ($\approx50$ per
cent) have absorption-line spectra. In addition to early type galaxies
however, these follow-up programs also showed that a significant
fraction of high-$z$ dust-shrouded active galaxies (starburst or AGNs)
are also present among EROs (Cimatti et al. 1998; Dey et al. 1999;
Afonso et al. 2001; Smith et al. 2001; Brusa et al. 2002).   

Although the relative mix between early type and dusty systems  
remains poorly constrained, EROs have strong spatial clustering which 
is comparable if not larger than that of present day luminous
ellipticals (Daddi et al. 2000, 2001, 2003; Firth et al. 2002; Roche
et  al. 2002, 2003). This may be interpreted as evidence that the
populations of distant EROs and nearby ellipticals may be evolutionary
linked. Study of the clustering of EROs, usually quantified via the
2-D correlation function, can indeed provide valuable information  on
the nature of these systems, their formation and evolution history as
well as their environment (e.g. Daddi et al. 2001; Roche et al. 2002, 
2003). Comparison of the clustering  properties of EROs with those of
local E/S0 may provide important insights into the evolution of
early-type galaxies out to $z\ga1$.

In this paper we explore the clustering properties and the environment
of EROs detected in a $\approx 180 \, \rm arcmin^{2}$ deep
($Ks\approx20$\,mag) $Ks$-band survey carried out as part of the
Phoenix Deep Survey (Hopkins et al. 2003), a large multiwavelength
program aiming to investigate the nature and the evolution of faint
(sub-mJy and $\mu$Jy) radio sources. Compared to previous studies of
the clustering of EROs, the present sample has the advantage of depth
combined with wider area providing a large sample of
EROs to $Ks\approx20$\,mag. This is essential to increase the sample
size and to improve the statistical reliability of the results
compared to previous studies at similar depths (e.g.  Roche et
al. 2002, 2003). We caution the reader however, that although our
survey is larger than previous samples at comparable magnitude limits,
cosmic variance is an issue and may affect our correlation amplitude
measurements. Much larger, degree scale, $K$-band surveys are required
to address this issue (e.g. NOAO Deep Wide Survey; Brown et
al. 2003). Our contiguous $Ks$-band survey  also allows study of the
association of EROs with regions of enhanced  galaxy density. Indeed,
although EROs are believed to reside in dense regions there is still
no direct link between EROs and galaxy overdensities. The overlap of
our sample with the  ultra-deep and homogeneous radio observations of
the Phoenix Deep Survey provides a unique opportunity to investigate
the association of the radio and the  ERO populations and how they
trace the underlying mass distribution.      

Section \S\,\ref{phoenix} presents the data used in this paper and
describes the selection of the ERO sample. The angular correlation 
function analysis is given in \S\,\ref{cor_fun}. 
\S\,\ref{density} outlines our analysis on the environment 
of EROs and \S\,\ref{radio} presents the cross-correlation of 
our ERO sample with the faint radio population.  Finally our results   
are discussed and our main conclusions are summarised in
\S\,\ref{discussion}.   
Throughout the paper we adopt $\rm 
H_{o}=70\,km\,s^{-1}\,Mpc^{-1}$, $\rm \Omega_{M}=0.3$ and  $\rm
\Omega_{\Lambda}=0.7$.  To allow comparison with previous 
studies all the distant dependent quantities are given in units of
$h=\rm H_{o}/100$.     

\section{The Phoenix Deep Survey}\label{phoenix}

\subsection{Radio and NIR/Optical data}

The Phoenix Deep Survey (PDS\footnote{\sf
  http://www.atnf.csiro.au/people/ahopkins/phoenix/}) is an on-going
survey studying the nature and the evolution of sub-mJy and $\rm
\mu$Jy radio galaxies. Full details of the existing radio, optical and
near-infrared data can be found in Hopkins et al. (2003) and Sullivan
et al. (2004); here we summarise the salient details. The radio
observations were carried out at the Australia Telescope Compact Array
(ATCA) at 1.4\,GHz during several campaigns between 1994 and 2001,
covering a 4.56 square degree area centered at RA(J2000)=$01^{\rm
  h}11^{\rm m}13^{\rm s}$ Dec.(J2000)=$-45\degr45\arcmin00\arcsec$. A
detailed description of the radio observations, data reduction and
source detection are discussed by Hopkins et al. (1998, 1999, 2003).
The observational strategy adopted resulted in a radio map that is
homogeneous within the central $\rm \approx1\,deg$ radius, with the
$1\sigma$ rms noise increasing from $\rm 12\mu Jy$ in the most
sensitive region to about $\rm 90\mu Jy$ close to the edge of the $\rm
4.56\,deg^2$ field. The radio source catalogue consists of a total of
2148 radio sources to a limit of 60\,$\rm \mu$Jy (Hopkins et al.
2003).

$Ks$-band near-infrared data of the central region of the PDS were
obtained using the SofI infrared instrument at the 3.6\,m ESO New
Technology Telescope (NTT).  The observational strategy and details of
the data reduction, calibration and source detection are described by
Sullivan et al. (2004). The $Ks$-band mosaic covers a $\rm 13.5 \times
13.3 \, arcmin^2$ area with a 45\,min integration time, and a central
$\rm 4.5 \times 4.5 \, arcmin^2$ subregion which has an effective
exposure time of 3\,h. The completeness limit is estimated to be
$Ks\approx20$\,mag for the full mosaic and $Ks\approx20.5$\,mag for
the deeper central subregion.

Deep multicolour imaging ($UBVRI$) of the PDF has been obtained using
the Wide Field Imager at the AAT on 2001 August 13 and 14
($BVRI$-bands) and the Mosaic-II camera on the CTIO-4m telescope on
2002 September 3 ($U$-band), fully overlapping the SofI $Ks$-band
survey. Full details on the data reduction, calibration and source
detection are again presented in Sullivan et al. (2004). In this study
we will use the $I$-band observations these being the deepest
($I\approx24.2$\,mag; see next section) and most appropriate for
identifying EROs.

\begin{figure}
\plotone{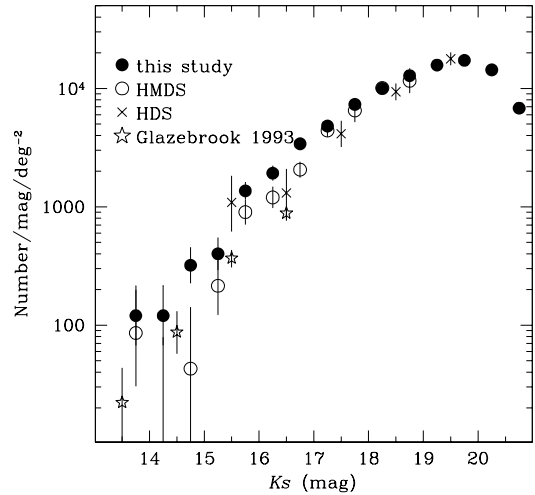}
\caption{
 $Ks$-band galaxy counts from the SofI survey  survey of the
 PDS (filled circles). Also shown are the $Ks$-band number counts from
 Glazebrook et al. (1994; stars) and from the Hawaii Deep (HDS;
 crosses) and  Medium  Deep (HMDS; open circles) surveys presented by
 Gardner, Cowie \& Wainscoat (1993).\label{fig_kcounts} 
 }
\end{figure}

\subsection{The ERO sample}\label{sample}

The $Ks$-band selected catalogue was constructed using {\sc
SExtractor} (Bertin \& Arnouts 1996). Sources were detected on the
$Ks$-band image and their photometric properties were then measured
from the seeing matched $Ks$ and the $I$-band frames (see Sullivan et
al. 2004 for details). To facilitate comparison with previous studies,
the estimated magnitudes are calibrated to the standard Vega based
magnitude system.   

The star-galaxy separation was performed using  the {\sc class\_star}
flag of {\sc SExtractor}  ({\sc class\_star}$>0.95$) which is reliable
to $Ks\approx16.5$. The PDS is selected to lie at high Galactic
latitude and therefore any contamination by stars at fainter
magnitudes is expected to be small. The $Ks$-galaxy number counts from
these observations are plotted in Figure \ref{fig_kcounts} showing
that the turnover magnitude lies at  $Ks\approx20$\,mag. Sullivan et
al. (2004) estimate moderate incompleteness of about 20--30 per cent
in the magnitude range $Ks=19.6-20.0$\,mag. In section \ref{cor_fun}
we argue that this small level of incompleteness is unlikely to alter
our main   conclusions. The central deep pointing of our $Ks$-band
survey has a turnover magnitude of $Ks\approx20.5$. In this paper
however, we consider sources brighter than $Ks=20$\,mag the
approximate completeness limit  of the entire catalogue.   

\begin{figure}
\plotone{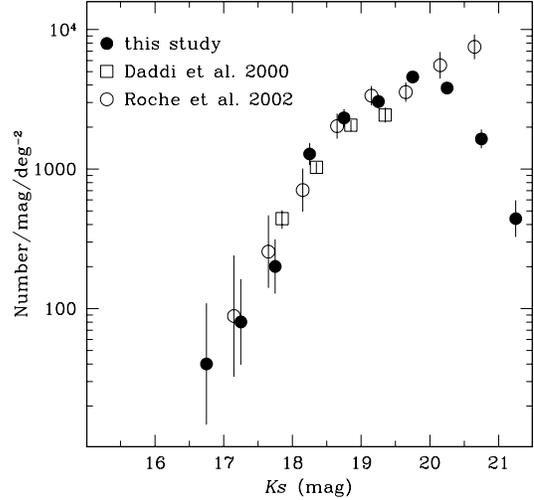}
\caption{
 $Ks$-band differential galaxy counts for the $I-K>4$\,mag EROs in the
 SofI $Ks$-band survey of the PDS (filled circles). Also shown are the
 $Ks$-band ERO counts from Daddi et al. (2000; squares) and Roche et
 al. (2002; open circles) offset for clarity by $+0.1$ and
 $-0.1$\,mag respectively. Our sample is complete to 
 $Ks\approx20$\,mag.\label{dnds_eros}
 }
\end{figure}

EROs are selected to have $I-K\ge4$\,mag with the $I$ and $Ks$-band
magnitudes measured in Kron (1980)-like elliptical apertures ({\sc
magauto} parameter of Sextractor). We note that the $5\sigma$
detection threshold for the $I$-band catalogue is 24.2\,mag,
sufficiently deep to identify  EROs with $Ks=20$\,mag. We find a
total of 289 EROs to this magnitude limit within the $\rm 13.5 \times
13.3 \, arcmin^2$ area of our survey. Figure 
\ref{dnds_eros} compares the differential counts of our EROs with
previous studies, suggesting that our sample is  complete to
$Ks\approx20$\,mag.   

A total of 95 radio sources brighter than $60\mu$Jy overlap with the 
$Ks$-band survey region. Using a matching radius of 3\,arcsec we find
that 17 of them with fluxes in the range $\rm 65-1000\mu Jy$ are
associated with EROs. The properties of these sources will be
presented in a future paper (Afonso et al. 2004, in preparation) and
are not considered in the analysis that follows. 

We also attempt to estimate the mean radio properties of EROs using
the stacking analysis method described by Hopkins et al. (2004). In
brief we extract sub-images from the radio mosaic at the location of
the non radio detected EROs, and construct  the weighted average of
the sub-images (weighted by 1/rms$^2$, to maximise the resulting
signal-to-noise, since the radio mosaic has a varying noise  level
over the image). Sub-images where low S/N emission ($>1.5\,\sigma$)
is present at the location of the non-detected source are excluded
from the stacking, in order to avoid biasing the stacking signal
result by the presence of a small number of low S/N sources. The
stacked image has an rms noise of $1.2\,\mu$Jy, and a
$\approx6.6\,\sigma$ detection at $8.0\,\mu$Jy. While the actual
redshift distribution of EROs remains poorly constrained, they are
believed to lie at $z\ga0.8$ (e.g. Cimatti et al. 2002; Cimatti et
al. 2003). Assuming that the average redshift of these sources is
$z\approx 1$, the inferred average ERO 1.4\,GHz luminosity is
$\rm 3.6 \times 10^{22}\, W \, Hz^{-1}$. For the luminosity estimate
we use a k-correction assuming a power law spectral energy
distribution of the form  $S_{\nu} \propto \nu^{-\alpha}$ with
$\alpha=0.8$. The luminosity above corresponds to an average star
formation rate (assuming the EROs are all star-forming systems) of
$\rm \approx 20 \, M_{\odot} \, yr^{-1}$, adopting the recent
calibration of Bell (2003). Clearly these results rely on a large
number of assumptions, but they serve the useful purpose of providing
a preliminary estimate for the average radio properties of 
these systems.

\section{Two point correlation function of EROs}\label{cor_fun}

The two-point angular correlation function, $w(\theta)$, is defined as the
joint  probability $\delta P$ of finding sources within the solid angle
elements $\delta \Omega_{1}$ and  $\delta \Omega_{2}$, separated by an
angle~$\theta$, in the form
 
\begin{equation}\label{eq_1}   
 \delta P= N^{2}\, (1+w(\theta)) \delta \Omega_{1} \delta \Omega_{2},
\end{equation} 

\noindent where $N$ is the mean surface density of galaxies. For a random
distribution of sources $w(\theta)$=0. Therefore, the  angular correlation
function provides a measure of galaxy density excess over that expected for
a random distribution. Various methods for estimating $w(\theta)$ have been
introduced, as  discussed by Infante (1994). In the present study, a source
is taken as the `centre' and the number of pairs within annular rings is 
counted. To account for the edge effects, Monte Carlo techniques are used by
placing random points within the area of the survey. We use the
 $w(\theta)$ estimator introduced  by Landy \& Szalay (1993) 
 
\begin{equation}
w(\theta)=\frac{DD-2DR+RR}{RR},
\end{equation}  

\noindent where DD and DR are respectively the number of sources and random
points at separations $\theta$ and  $\theta + d\theta$ from a given
galaxy. Similarly, RR is the number of random points within the
angular interval $\theta$ to $\theta + d\theta$ from a given random
point. 
 
For a given ERO sample, a total of 100 random catalogues are
generated, each having the same number of points as the original data
set. The random sets were cross-correlated with the galaxy catalogue,
giving  an average value for $w(\theta)$ at each angular
separation. The uncertainty in $w$($\theta$) is determined from the
relation 

\begin{equation}
\sigma_{w}=\sqrt{\frac{1+w(\theta)}{DD}}.
\end{equation}  

\begin{figure}
\plotone{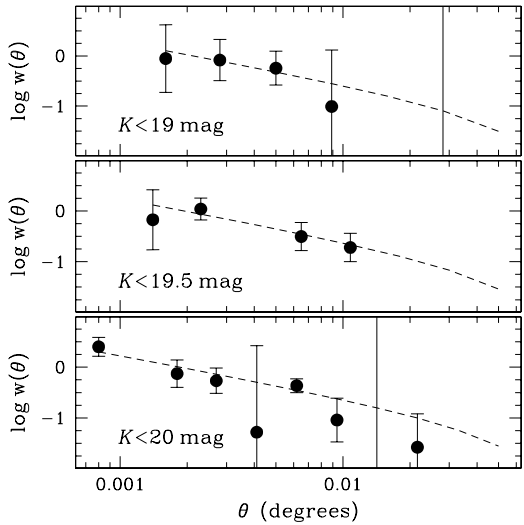}
\caption{
 Angular correlation function $w(\theta)$ for the three
 magnitude limited subsamples listed in Table \ref{tab_1}. The lines
 are the best fit to the observations.\label{fig_corfun} 
 }
\end{figure}

\begin{figure}
\plotone{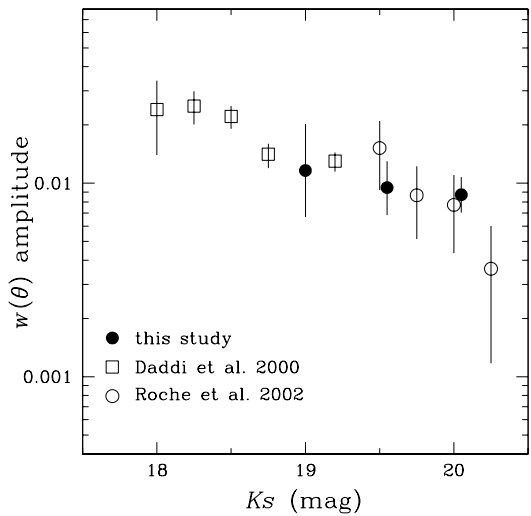}
\caption
 {
The amplitude of the angular correlation function of EROs as a
function of the limiting $Ks$-band magnitude. The filled circles are
the results from this study, after applying the correction for stellar
contamination, the squares and the open circles are the amplitudes
from Daddi et al. (2000) and Roche et al. (2002) 
respectively.\label{fig_aw}
 }
\end{figure}

\noindent Finally, before fitting a power law to $w(\theta)$, we take
into account a bias arising from the finite boundary  of the
sample. Since the angular correlation function is calculated within a
region of solid angle $\Omega$,  the background projected density of
sources, at a given magnitude limit, is effectively 
$N_{s}/\Omega$ (where $N_{s}$ is the number of sources  brighter than
the limiting magnitude). However, this is an overestimation of the true
underlying mean surface density, because of the positive correlation
between  galaxies in small separations, balanced by negative values of
$w(\theta)$ at larger separations. This bias, known  as the integral
constraint, has the effect of reducing the amplitude  of the
correlation function by

\begin{equation}\label{eq_8}
\omega_{\Omega}=\frac{1}{\Omega ^{2}} \int{\int{w(\theta) d\Omega_{1} d\Omega_{2}}},  
\end{equation} 
 
\noindent where $\Omega$ is the solid angle of the survey
area. Following Roche et al. (2002) we estimate $\omega_{\Omega}$
numerically using the random-random correlation

\begin{equation}\label{eq_9}
\omega_{\Omega}=\frac{\sum{RR \, A_w \, \theta^{-\delta}}}{\sum{RR}},  
\end{equation} 

\noindent
where we assume that $w(\theta)$ is a power-law of the form  $A_w \,
\theta^{-\delta}$. The sum in equation \ref{eq_9} is from 1\,arsec
to 15\,arcmin. Adopting $\delta=0.8$ we find $\omega_{\Omega}=6.9
\times A_w$. Having fixed the exponent $\delta$ to 0.8 the amplitude
$A_w$ is obtained by fitting the function $A_w \theta^{-0.8} -
\omega_{\Omega}$ to the observed $w(\theta)$  using standard $\chi^2$
minimisation procedures weighting each  point with  its error. 
We estimate the $w(\theta)$ and determine the amplitude $A_w$ for
different $Ks$-band magnitude limited ERO subsamples. The results are
summarised in Table \ref{tab_1}. The estimated correlation function is
plotted in Figure \ref{fig_corfun}.  For the $Ks<19$ and $<19.5$\,mag
sub-samples the correlation function signal is significant at the
$2\sigma$ level with $w(\theta<25^{\prime\prime})= 0.68\pm 0.30$ and
$0.28\pm 0.13$ respectively. For the sub-sample with $Ks<20$\,mag the   
detected signal is significant at the $3.7\sigma$ confidence level
($w(\theta<25^{\prime\prime})= 0.37\pm 0.10$). The  amplitudes
estimated above must be corrected for contamination of the  galaxy
sample by  stars at faint magnitudes ($Ks>16.5$\,mag). The presence of
an uncorrelated population of uniformly distributed  stars within the
galaxy catalogue, reduces  $A_{w}$ by the factor   

\begin{equation}\label{eq_3.19}
 \biggl(\frac{N_{obj}}{N_{obj}-N_{s}}\biggr)^{2},
\end{equation}

\noindent where $N_{obj}$ is the number of objects (both stars and
galaxies) used to calculate $w(\theta)$ and $N_{s}$ is the number of stars
in the catalogue. In the present study, stars were identified and removed
to $K=16.5$\,mag (section 2.2). At fainter magnitudes,
the Milky Way stellar population synthesis model described by
Robin et al. (2003, 2004) was employed to predict the expected number
of stars within our ERO sample at different $Ks$-band magnitude
limits. The correction factors estimated from equation
\ref{eq_3.19} are also listed Table \ref{tab_1} along with the
corrected amplitudes, $A_w^s$. The $A_w^s$  (with $\theta$ measured in
degrees) are consistent with previous studies at similar magnitude
limits (e.g. Daddi et al. 2001; Roche et al. 2002, 2003). This is
demonstrated in Figure \ref{fig_aw} plotting the $A_w^s$ for EROs as a
function of the limiting $Ks$-band magnitude. Compared to previous ERO
surveys to $Ks=20$\,mag our sample has the advantage of larger size
and hence more statistically reliable  results. Cosmic variance
however, is likely to be an issue for surveys with  areal extent
similar to our own. Daddi et al. (2001) estimated that the relative
dispersion on the amplitude of the angular correlation function due to 
this effect is $\sigma_{A_w} / A_w =  \sqrt{A_w\,\,\omega_{\Omega}}$,
where  $\omega_{\Omega}$ is the integral constraint defined in
equation \ref{eq_8}.  For our survey geometry using this relation we
estimate an uncertainty to the amplitudes listed in Table
\ref{tab_1} due to cosmic  variance of $\approx20\%$. We note however,
that this result is likely to be a lower limit since it does not take
into account the higher order moments of the galaxy
distribution. For example Barger et al. (1999) and McCracker et
al. (2000) find a factor of three difference in the surface density of
EROs to $K\approx20$\,mag in their independent  small area surveys
($\rm \approx 60\, arcmin^2$). Although our sample has 3 times larger
field of view  the result above illustrates the effect of cosmic
variance to relatively small area surveys. Much larger, degree scale,
$K$-band samples will be able to address this  issue (e.g. NOAO Deep
Wide Survey; Brown et al. 2003). N-body  simulations can also be used
in principle to provide  a realistic estimate of the cosmic variance
effect (e.g. Somerville et al. 2004). This calculation however,
requires a number of assumptions on the redshift distribution and
clustering properties of EROs and is beyond the scope of this paper.  

We also attempt to quantify the effect of the $Ks$-band catalogue 
incompleteness at faint magnitudes (20-30\% in the range 19.6-20\,mag)
to the estimated $w(\theta)$ of EROs. We need a sample that does not
suffer from incompleteness in the above magnitude range to create
mock catalogues by randomly removing 20-30\% of the sources at faint
magnitudes. The angular correlation  function for each mock catalogue
is then estimated to quantify the bias introduced by the $Ks$-band
catalogue incompleteness.  The above prescription could have been
applied to  the central $\rm  4.5 \times 4.5 \, arcmin^2$ subregion of
our survey which is deeper in the $Ks$-band ($Ks\approx20.5$\,mag) and
thus does not suffer from  incompleteness in the 19.6-20\,mag
range. However, only about 30 EROs lie in this subregion and therefore
the small number statistics do not allow us to use this
sample. Nevertheless, we apply the above method to the $Ks<19.5$\,mag 
subsample (total of 177 sources, see Table \ref{tab_1}) by randomly
removing 20-30\% of the sources in the range 19.0-19.5\,mag. A total
of 500 mock catalogues are created and the amplitude of the angular 
correlation function of each one of them is estimated. Within the
$1\sigma$ uncertainties we find no difference between the mean $A_w$
from the 500 mock realisations and that quoted in Table
\ref{tab_1}. Extrapolating the above result to the $Ks<20$\,mag we 
argue that the estimated amplitude is not going to be significantly
affected by incompleteness at faint magnitudes. We caution the reader
however, that the above conclusion holds only if the missed objects
have the same clustering properties as the detected sources.

The agreement between the angular correlation amplitudes estimated
here and those found in previous studies also implies an agreement on
the spatial correlation length $r_{o}$ assuming a similar redshift
distribution, which is not unreasonable given the similarity of the
sample selection. Use of the luminosity function models of Daddi et
al. (2001) and Roche et al. (2002, 2003) to deproject the ERO
angular correlation amplitudes and to determine their spatial
correlation  length, $r_{o}$, yields  $r_{o}=12-17\,h^{-1}  \rm 
\,Mpc$. The range in  $r_{o}$ depends on the adopted luminosity
function and clustering evolution model. As discussed above the
uncertainty in $r_{o}$ due to cosmic variance is estimated to be at
least $\approx20\%$. We also caution the reader that the
relatively small extent of our survey may result in systematic
underestimation of the derived correlation length. For example, Daddi
et al. (2001) estimate this effect to be $<10\%$.

\begin{table} 
\footnotesize 
\caption{
	\rm Clustering amplitudes of the  $I-K>\rm 4\,mag$ EROs at
	different magnitude limits. The $A_w$ is estimated at
	1\,degree and is {\it not} corrected for dilution due to
 	stellar contamination of the 
	ERO sample at faint magnitudes. The correction factors in the last   
	column account for this effect providing an estimate of the reduction
	of the measured $A_w$ due to the presence of stars (see text for
	details). The last column gives the clustering amplitudes, $A^s_w$,
	corrected for stellar contamination 
	}\label{tab_1}  
\begin{center} 
\begin{tabular}{ccccc} 
\hline
  $Ks$-band       &Number of& $A_w$               & correction & $A^s_w$ \\  
 limit (mag)      &  EROs   & ($\times 10^{-3}$)  & factor    & ($\times 10^{-3}$) \\
\hline
 19.0             &  100   & $7.7^{+5.6}_{-3.3}$ & 1.52 & $11.7^{+8.5}_{-5.0}$\\  
 19.5             &  177   & $7.1^{+2.6}_{-2.0}$ & 1.34 & $9.5^{+3.5}_{-2.7}$\\  
 20.0             &  289   & $6.8^{+1.6}_{-1.3}$ & 1.28 & $8.7^{+2.1}_{-1.7}$\\  
\hline
\end{tabular} 
\end{center} 
\normalsize  
\end{table} 

\section{The environment of EROs}\label{density}

In this section we study the environment of EROs to explore their
association with regions of enhanced density at relatively high
redshifts ($z\approx1$). We quantify this by cross-correlating the
positions of our ERO sample with the smoothed $Ks$-band galaxy density
map produced using the matched filter algorithm described by
Postman et al. (1996). 

\subsection{Creating the density map}\label{mfa}

The matched filter algorithm was developed by Postman et
al. (1996) to identify galaxy overdensities using photometric data
only. It has the advantage that it  exploits both  positional and
photometric information producing galaxy density maps where spurious
galaxy fluctuations are suppressed. A drawback of  the matched filter
method is that one must assume a form for the cluster luminosity
function and its radial profile. Clusters with the same richness but
different intrinsic shape or different luminosity function from the
adopted ones do not have the same likelihood of being  detected. 

A detailed description of the matched filter algorithm can be found in
Postman et al. (1996). In brief, the galaxy catalogue is convolved
with a filter derived from an approximate maximum likelihood
estimator obtained from a model of the spatial and luminosity
distribution of galaxies within a cluster. The luminosity weighting
function (i.e. flux filter) is defined as

\begin{equation}\label{eq_5}
L(m)=\frac{\phi(m-m^{*})\,10^{-0.4\,(m-m^*)}}{b(m)},
\end{equation}

\noindent 
where $\phi(m-m^{*})$ is the cluster luminosity function, $m^{*}$ is
the apparent magnitude corresponding to the characteristic luminosity
of the cluster luminosity function and $b(m)$ is the surface density
of field galaxies with apparent magnitude $m$. We adopt a Schechter
form for the cluster $Ks$-band luminosity function with parameters 
$\alpha=-1.09$ and $M^{*}=-23.53+5{\rm log}h$ estimated by Kochanek et
al.  (2001) for early type galaxies. The term $10^{-0.4\,(m-m^*)}$ in
equation \ref{eq_5} is introduced to avoid divergence of the integral
of $L(m)$ at faint magnitudes in the case of  Schechter luminosity
functions with $\alpha<-1$. The spatial weighting function
(i.e. radial filter) is assumed to follow the form   
 
$$
P(r) =\left\{
\begin{array}{cc}
\frac{1}{\sqrt{1+(r/r_c)^2}}- \frac{1}{\sqrt{1+(r_{co}/r_{c})^2}}, &\mathrm{if \,\,\,r<r_{co}} \nonumber \\
0 & \mathrm{otherwise,} \\
\end{array}
\right.
$$


\noindent 
where $r_c$ is the cluster core radius and $r_{co}$ is an
arbitrary cutoff radius. Here, we adopt $r_{c}=100\,h^{-1}$\,kpc 
and $r_{co}=1\,h^{-1}$\,Mpc (e.g. Postman et al. 1996). In practice
the survey area is binned into pixels $(i,j)$ of fixed size (for the
choice of values see below) and the observed galaxy distribution is
convolved with the filters above producing  a likelihood map according
to the relation  

\begin{equation}\label{eq_3}
S(i,j)=\sum_{k=1}^{N_T}\,P(r_k)\,L(m_k),
\end{equation}

\noindent 
where $N_T$ is the total number of galaxies in the catalogue. The
sum above is evaluated for every pixel of the density map. Both
$m^{*}$ and $r_{c}$ are a function of redshift and hence $S(i,j)$ also
depends on redshift through these parameters. The redshift dependence
of $m^{*}$ also includes a $k$-correction. Here we adopt a
non-evolving elliptical galaxy model obtained from the Bruzual \&
Charlot (1993) stellar population synthesis code as described in
Pozzeti, Bruzual \& Zamorani (1996).   

We apply the matched filter algorithm to galaxies brighter than $Ks =
20$\,mag in our $Ks$-band survey to avoid biases due to incompleteness
at fainter magnitudes. The galaxy density map $S(i,j)$ is
independently estimated for redshifts between $z_{min}=0.5$ to
$z_{max}=1.2$, incremented in steps of 0.1. The $z_{max}$ corresponds
to the redshift where  $m^{*}$ becomes comparable to the limiting
magnitude of the survey.  The characteristic luminosity, $L^{*}$, the
faint end slope of the luminosity function, $\alpha$, and the cluster
core radius, $r_{c}$, are assumed to remain constant with
redshift. Adopting a passively evolving $L^{*}$ for the luminosity
function does not alter our main conclusions but, as discussed below,
only shifts $z_{max}$ to higher redshifts, $z_{max}\approx1.6$. For
the passive evolution we adopt the elliptical galaxy model  described
by  Pozzeti, Bruzual \& Zamorani (1996) which predicts a brightening 
of $M^{*}$ by about 0.8 magnitudes to $z\approx1$. This is in fair
agreement with recent studies on the $K$-band luminosity function
evolution (Pozzetti et al. 2003; Toft et al. 2004; Ellis \& Jones
2004). The results presented here assume a constant $L^{*}$ with
redshift. The pixel size of the galaxy density maps at any redshift is
taken to be $\approx18$\,arcsec corresponding to a projected cluster
core radius of $r_{c}=100\,h^{-1}$\,kpc at the redshift $z=1$. Figure 
\ref{dense_map} shows the $z=1.1$ density map generated by the method
above with the positions of EROs overplotted.      
 
\begin{figure} 
\vspace{7cm}
 \caption {
  $Ks$-band density map with parameters tuned to $z=1.1$. The contours
  delineate the regions with overdensity $>1$ (i.e. above the mean
  density of the map). The circles show the positions of $Ks<20$\,mag
  EROs in our sample.  
  }\label{dense_map}  
\end{figure}

\subsection{ERO/density-map cross-correlation}\label{mfa_res}

The cross-correlation function between the $Ks<20$\,mag EROs and the
$Ks$-band density map (at a given redshift) is estimated using the
relation

\begin{equation}
w_{E,D}(\theta) = \frac{1}{N_{eros}} \, \sum_{k}^{N_{eros}}\sum_{i,j}(S_{i,j} - \bar{S}),
\end{equation}

\noindent
where $N_{eros}$ is the total number of EROs, $S_{i,j}$ is the value
of the $(i,j)$ pixel of the density map produced using the
prescription described in the previous section, $\bar{S}$ is the mean
value of the density map and the sum is for all pixels with angular
separation $\theta$ from a given ERO. The uncertainties were estimated
from 100 bootstrap resamples of the EROs. Simulated data sets were
generated by sampling N points with replacement from the true ERO
dataset of N points. The cross-correlation function is then estimated
for each of the bootstrap samples in the same fashion as with the real
dataset. The standard deviation around the mean for a given angular
separation $\theta$ is the used to estimate the uncertainty in the
$w_{E,D}(\theta)$. 

To assess the expectation in the case of a random distribution of
points we produce 100 mock $Ks$-band catalogues by randomising the 
positions of the $Ks$-band galaxies. We then apply the matched filter
algorithm to produce  density maps for each of the independent mock
catalogues. The (randomised) positions of the EROs in the mock
catalogues are then cross-correlated  with the density maps in the
same fashion as for the real dataset.  For each separation
$\theta$ the above procedure provides an estimate of the
cross-correlation function  expected in the case of a random
distribution of galaxies (i.e. without clustering). This procedure
also takes into  account the presence of spurious galaxy overdensities
that may be produced by the matched filter algorithm. 

\begin{figure} 
 \plotone{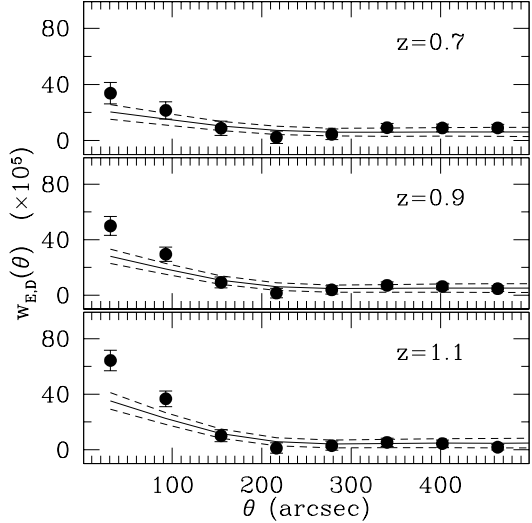} 
 \caption {
 Cross-correlation function $w_{E,D}(\theta)$ of the
 $Ks<20$\,mag EROs and the density map produced by the matched filter
 algorithm of Postman et al. (1996) at redshifts $z=0.7$, 0.9 and
 1.1. The continuous line is the expectation for a random distribution
 of galaxies and the dashed lines are the $1\sigma$ uncertainty
 envelop  around the mean.\label{fig_cross_combo}
 }
\end{figure}

\begin{figure} 
 \plotone{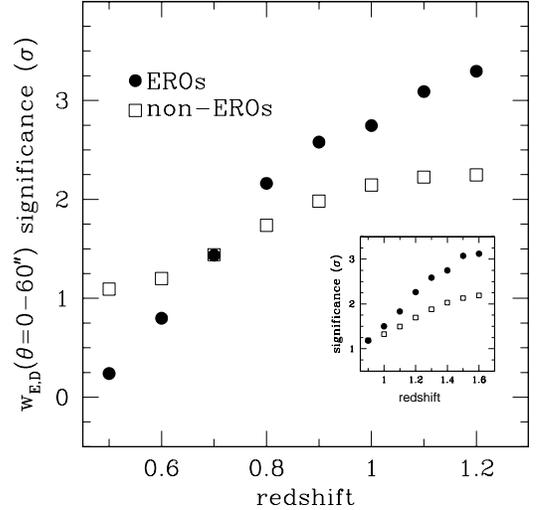} 
 \caption {
 Significance above the random expectation of the  detected
 $w_{E,D}(\theta)$ signal for separations 0--60\,arcsec as a function
 of the redshift that the density map (used in the cross-correlation)
 was estimated assuming an non-evolving $L_{*}$. The filled circles
 are the $Ks<20$\,mag ERO sample  
 ($I - K>4$\,mag) and the open squares are non-ERO (i.e. $I-K<4$)
 $Ks$-band selected galaxies (see text for details). The significance
 is in standard deviations, $\sigma$, taking into  account the
 uncertainties in both the $w_{E,D}(\theta)$ and the random
 expectation. The insert plot shows the same results
 ($w_{E,D}(\theta)$ significance versus $z$) in the case of density
 maps estimated assuming passively evolving $L_{*}$ luminosity. There
 is no qualitative difference between the non-evolving and passively
 evolving $L_{*}$ results except from the $z_{max}$ which shifts from
 1.2 to 1.6 in latter case.  
 }\label{fig_cross_sign}   
\end{figure}

Using the procedure above we cross-correlate the $Ks<20$\,mag EROs with
the density maps produced by the matched filter algorithm with
parameters tuned at  redshifts $z=0.5-1.2$. The results for the
$z=0.7$, 0.9 and 1.1 density maps are plotted in Figure
\ref{fig_cross_combo} along with the random expectation. There is
evidence for a statistically significant ($\approx3\sigma$) positive
signal for angular separations 0--60\,arcsec in the case of the high
redshift density maps (e.g. $z=0.9$, 1.1).  At lower redshifts
(e.g. $z=0.7$) we find no signal above the random expectation. This is
further demonstrated in Figure \ref{fig_cross_sign} plotting the
significance above the random expectation of the  detected
$w_{E,D}(\theta)$ signal for separations 0--60\,arcsec (i.e. the first
bin in Figure \ref{fig_cross_combo}) as a function of the redshift
that the density map was estimated.  The significance in this figure
is expressed in units of standard deviations, $\sigma$, taking into
account the uncertainties in both the $w_{E,D}(\theta)$ and the random 
expectation. In Figure  \ref{fig_cross_sign} the significance of the
$w_{E,D}(\theta)$ signal increases with redshift to about $3\sigma$
when the cross-correlation is performed with density maps generated for
redshifts  $z\ga1$. Adopting a passively evolving $L^{*}$ for the
luminosity function (see section \ref{mfa}) shifts the curve
in Figure \ref{fig_cross_sign} to higher redshifts with the peak at
$z=1.2$ moving to $z=1.6$ but does not qualitatively alter the
results. This is demonstrated in the insert plot of Figure
\ref{fig_cross_sign}. The evidence above suggests that EROs are
associated with  high-$z$ overdensities. 

We also explore if non-ERO (i.e. $I-K<4$) $Ks$-band selected galaxies
with the same magnitude distribution as the ERO sample give equally
significant cross-correlation signal. We produce a total of 100 mock
catalogues by randomly selecting galaxies from the full $Ks$-band
source list with $I-K<4$ (i.e. not EROs) and magnitude distribution
similar to that of our $Ks<20$\,mag ERO sample. Each of these mock
catalogues are cross-correlated with the $Ks$-band density maps
providing, at different redshifts, an estimate of the mean
cross-correlation function and its rms for each separation. This is
then compared against the signal for EROs. For the $z=1.2$ and $1.1$
maps we find that the ERO signal is higher than that of non-ERO
galaxies, on average, at the $2.6\sigma$ and $2.1\sigma$ confidence
level respectively. For lower redshifts the EROs have
cross-correlation signal consistent with that of  non-EROs within the
$1\sigma$ uncertainties. This is demonstrated in Figure
\ref{fig_cross_sign} plotting the significance above the random 
expectation of the mean $w_{E,D}(\theta)$ signal for non-ERO $Ks$-band
selected galaxies for separations 0--60\,arcsec as a function of the
redshift that the density map was estimated. Together this all
provides direct evidence that  EROs are associated with regions of
enhanced density at $z\ga1$.  

\section{EROs and the faint radio population}\label{radio}

Radio galaxies are another class of sources that are believed to be
good cluster tracers to high-$z$ (e.g. Zirbel 1997; 
Zanichelli et al. 2001). We exploit the deep 1.4\,GHz radio 
observations available for our $Ks$-band survey to explore the
association of EROs with radio sources and whether they trace similar
structures. These are  quantified using the two-point
cross-correlation function $w_{1.4,E}(\theta)$ between radio sources
and EROs.

\begin{figure} 
 \plotone{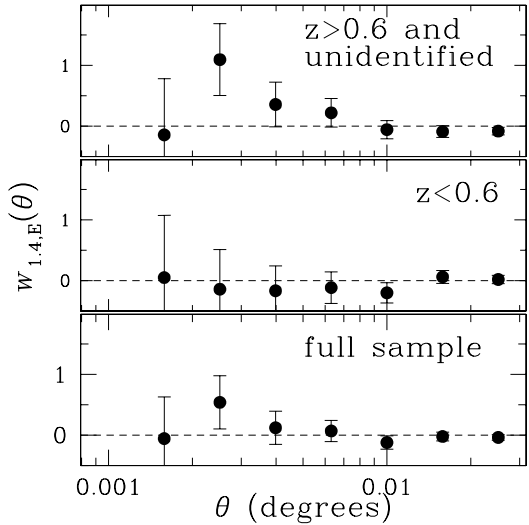} 
 \caption {
 Cross-correlation function $w_{1.4,E}(\theta)$ of the faint radio
 sources with the $Ks<20$\,mag EROs. The dashed lines in all panels
 mark the expectation for a random distribution. The 
 bottom panel is the $w_{1.4,E}(\theta)$ using the full sample (78) 
 of faint radio sources. The middle panel is the cross-correlation
 function using the low-$z$ sub-sample with photometric redshift
 estimates $z<0.6$. The top panel is for the faint radio source
 sub-sample with photometric redshifts $z>0.6$ or no optical
 counterparts.\label{fig_corfun_radio_eros}  
 }
\end{figure}

The cross-correlation function is estimated taking radio sources as
centres and counting the number of EROs around  them in successive
annuli. This is then compared with the expectation for a random
distribution by placing  a total of 40\,000 random points in the
surveyed area and counting the number of pairs between radio sources
and the random points. The cross-correlation function is  defined  

\begin{equation}
w_{1.4,E}(\theta)=\frac{DD(1.4,E)\,\,N_{R}}{DR(1.4,R)\,\,N_{E}} - 1,
\end{equation}  

\noindent 
where $DD(1.4,E)$,  $DR(1.4,R)$ is the number of radio/ERO and
radio/random pairs respectively with separation $\theta$ and $N_{R}$,
$N_{E}$ is the total number of random points and EROs respectively. As
already discussed, in the cross-correlation we have excluded a total
of 17 radio sources that  are associated with EROs. The results for
the remaining 78 radio sources in our sample are plotted in the bottom
panel of Figure \ref{fig_corfun_radio_eros} with the errors being
Poisson estimates. There is little signal with the estimated
$w_{1.4,E}(\theta)$  consistent with zero. This suggests that the bulk
of the EROs and the radio source populations are probing different
structures.    

We further explore this by grouping the radio sources into high and
low-$z$ sub-samples using the photometric redshift estimates presented
by Sullivan et al. (2004). The low-$z$ sub-sample comprises 35 sources
with $z_{phot}<0.6$  while, the  high-$z$ sub-sample includes a
total of 43 sources both without optical counterparts to the limit  
$I\approx24$\,mag (assumed to lie at  high redshifts) and with
$z_{phot}>0.6$. The cross-correlation results using the above
sub-samples are plotted in the middle (low-$z$) and top panels
(high-$z$) of Figure \ref{fig_corfun_radio_eros}. Although there is no
signal in the case of the low-$z$ radio sources, the high-$z$
sub-sample shows signal for separations 7--30\,arcsec, albeit at the
$\approx2\sigma$ significance level ($w_{1.4,E}(7-30^{\prime\prime})=
0.34\pm0.18$). Although the statistics are poor the evidence above
suggests that the different redshift distributions of the radio and
ERO populations are responsible for the lack of cross-correlation
signal in the bottom panel of  Figure \ref{fig_corfun_radio_eros}. The
EROs peak at $z\ga1$ (e.g. Cimatti et al. 2001, 2003; Firth et al. 
2002; V\"{a}is\"{a}nen \& Johansson 2004a) while many of the radio
sources to the flux density limit of the PDS lie at $z\la1$
(Georgakakis et al. 1999; Afonso et al. 2004; Sullivan  et al. 2004).    

This is also supported by the cross-correlation of the full radio
sample with the $Ks$-band density maps produced in section \ref{mfa}.
The cross-correlation is performed using the method outlined in
section \ref{mfa_res}. The expectation in the case of a random
distribution of points is quantified by producing 100 mock radio
catalogues  by randomising the positions of the radio sources. Each of
these random catalogues is cross-correlated with the $Ks$-band density
maps providing,  for a given separation $\theta$, an estimate of the 
cross-correlation function  expected in the case of a random
distribution of galaxies (i.e. without clustering).

\begin{figure} 
 \plotone{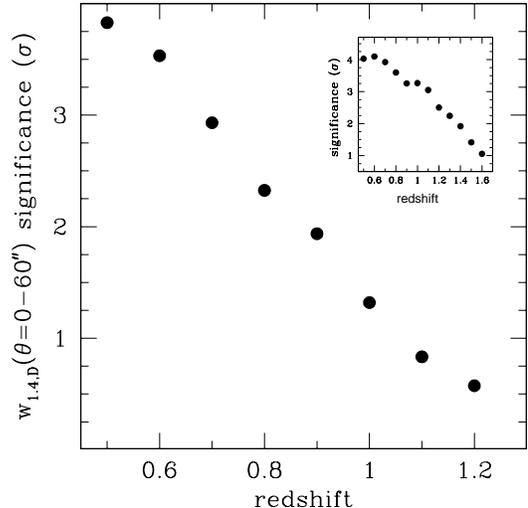} 
 \caption {
 Significance above the random expectation of the  cross-correlation 
 function, $w_{1.4,D}(\theta)$,  between the faint radio
 sources and the $Ks$-band  density maps at different redshifts
 produced as described in section  \ref{density}. The density maps are
 estimated assuming a non-evolving $L_{*}$ for the luminosity
 function. We plot the $w_{1.4,D}(\theta)$ signal for separations
 0--60\,arcsec expressed in standard deviations, $\sigma$, taking into
 account the uncertainties in both the $w_{1.4,D}(\theta)$ and the
 random expectation.The insert plot shows the same results
 ($w_{1.4,D}(\theta)$ significance versus $z$) in the case of density
 maps estimated assuming passively evolving $L_{*}$ luminosity. There
 is no qualitative difference between the non-evolving and passively
 evolving $L_{*}$ results. 
 }\label{fig_cross_sign_radio} 
\end{figure}

The results are shown in Figure \ref{fig_cross_sign_radio} plotting
the significance above the random expectation of the  detected 
cross-correlation signal for separations 0--60\,arcsec as a function
of the redshift that the density map was estimated (assuming a
non-evolving $L^{*}$).  The significance in this figure is expressed
in units of standard deviations, $\sigma$. In Figure
\ref{fig_cross_sign_radio} the significance of the cross-correlation
signal decreases with redshift falling below the $\approx3\sigma$ when
the cross-correlation is performed with density maps generated for
redshifts  $z>0.7$. Adopting a passively evolving $L^{*}$ for the
luminosity function (see section \ref{mfa}) does not qualitatively
change our results but only shifts to higher-$z$ the point that the
significance of the  cross-correlation signal falls below
$\approx3\sigma$. This is demonstrated in the insert plot of Figure  
\ref{fig_cross_sign_radio}. 

The trend for radio sources in Figure \ref{fig_cross_sign_radio} is
opposite to that for EROs in Figure \ref{fig_cross_sign} suggesting
that the bulk of the radio and the  ERO populations are tracing
structures at different redshifts.    

\section{Discussion and summary}\label{discussion}
In this paper we explore the clustering properties of $Ks<20$\,mag EROs
($I-K>4$\,mag) using a  $Ks$-band sample covering $\rm \approx 180\,
arcmin^2$ and overlapping with the PDS region. We use angular
correlation function analysis to estimate a statistically significant
clustering signal ($>3\sigma$) with  amplitude consistent, within the
errors, with previous studies (Roche   et al. 2002, 2003; Daddi et
al. 2000). The advantage of our study is that our sample size is
larger than previous surveys  to $Ks=20$\,mag (e.g. Roche et al. 2002,
2003) providing better statistical reliability. Nevertheless, cosmic
variance remains an issue in our survey. The $w(\theta)$ amplitudes
estimated here  are consistent with 3-D clustering lengths  $r_{o}=\rm
12-17\, h^{-1}\,Mpc$ depending on the adopted model luminosity
function and the evolution of the clustering (Daddi et al. 2001; Roche
et al. 2002, 2003). These $r_{o}$ values are in-between those of 
present-day   ellipticals (e.g. Guzzo et al. 1997) and Abell clusters
(Abadi, Lambas  \& Muriel 1998) and comparable to those estimated for
hard X-ray selected sources (Basilakos et al. 2004) and SCUBA sources
(Almaini et al. 2003). Daddi et al. (2002) also discuss evidence that
EROs with optical/NIR  colours suggesting old stellar population are
significantly more clustered than those showing evidence for dusty
starbursts. 

A number of studies also suggest a dependence of the correlation
length on the galaxy absolute luminosity (e.g. Willmer, da Costa \&
Pellegrini 1998; Norberg et al. 2002; Zehavi 2002; Brown et
al. 2003). For example, Brown et al. (2003) estimated the correlation
length of $z<1$ red ($B_W - R > 1.44$) galaxies in the NOAO Deep Wide
Field Survey and found that systems with $L>L^{*}$ have higher
clustering lengths ($r_{o} \approx 10 \, h^{-1} \rm \, Mpc$) than
$<L^{*}$ galaxies ($r_{o} \approx 5 \, h^{-1} \rm \, Mpc$). Similar
results are obtained for the early type galaxies in the 2dF Galaxy
Redshit Survey (Norberg et al. 2002). The EROs identified in our
sample comprise systems with $L \la L^{*}$. For example at $z=1$ and
1.5  the magnitude limit  $Ks=20$\,mag corresponds to about
$0.2\,L^{*}$  and $0.5\,L^{*}$ respectively, assuming a passively
evolving elliptical galaxy model. The large clustering lengths of 
$Ks\approx20$\,mag EROs  ($r_{o}=12-17\,  h^{-1} \rm \,Mpc$) are
therefore difficult to reconcile with the above $r_{o}$ estimates for
$L \la L^{*}$ early type galaxies at $z<1$. This may be due to
uncertainties in the model redshift distribution of EROs used to
deproject the angular correlation function amplitudes. Combination of
photometric and spectroscopic redshifts for complete ERO samples are
essential to refine these models and to better understand the
clustering properties of this population. 

The evidence above on the large correlation length of EROs may
suggest these sources are tracing high density regions. We further
explore the association of EROs with galaxy overdensities using a
novel method: we first smooth the $Ks$-band galaxy distribution using
the matched filter algorithm of Postman et al. (1996) and then
cross-correlate the resulting density maps with the ERO positions. The
matched filter algorithm parameters are tuned to produce density maps
each one of which is sensitive to galaxy clusters at a different
redshift in  the range 0.5--1.2. We find a statistically significant
cross-correlation signal ($\ga3\sigma$ compared to the random
distribution) only for density maps tuned to $z\ga1$ clusters. At
these redshifts, EROs also have  higher cross-correlation signal,
albeit at the  $2-2.5\sigma$ confidence level, than the non-ERO
$Ks$-band selected galaxies  with the same magnitude
distribution. This provides direct evidence that EROs are, on average,
associated with regions of enhanced galaxy density at redshifts
$z\ga1$. Previous studies also claim that EROs lie within rich cluster
regions at high-$z$.  Daddi et al. (2000) and Roche et al. (2002) have
identified  ERO overdensities within their surveys and argued that
these may be associated with massive clusters at
$z\ga1$. V\"{a}is\"{a}nen  \& Johansson (2004b) also report
overdesnities of EROs in the vicinity of faint mid-IR  ISO sources and
argue that these may be associated with high-$z$  clusters. Similar
enhanced ERO number densities have been reported in high-$z$ AGN and
QSO fields (e.g. McCarthy, Presson \& West 1992; Hu \& Ridgeway 1994;
Chapman, McCarthy \& Persson 2000; Hall et al. 2001; Best et al. 2003; 
Wold et al. 2003), although is not yet clear if these ERO
overdensities are indeed associated with clusters linked to  the
central AGN. Our analysis provides direct evidence that EROs are
associated with high-$z$ overdense regions and confirms the  claims of
the studies above.

Also, our finding that EROs trace dense regions at $z\ga1$ is
consistent with previous studies on the redshift distribution 
of EROs. The red $I-K$ colours of these systems can only be 
explained by either early type galaxies or dusty starbursts  at 
$z\ga0.8$ (e.g. Pozzetti \& Mannucci 2000; V\"{a}is\"{a}nene \&
Johansson 2004a; Roche et al. 2002, 2003). Luminosity function models
used to interpret the observed ERO counts and their clustering
properties produce spiky redshift distributions  with a peak at
$z\approx1$ and a long tail extending to high-$z$ (Daddi et al. 2001;
Roche et al. 2003). Spectroscopic and photometric redshifts for EROs
also show that most of them are associated with $z\ga1$ ellipticals or
dusty starbursts, although the relative mix of the two populations
remain poorly constrained (e.g.  Cimatti et al. 1998; Dey 1999; Piere
et al. 2001; Smith et al. 2001; Afonso et al. 2001; Cimatti et
al. 2001;  Daddi et al. 2003; Martini 2001). 

We also investigate the association between EROs and faint radio
sources and how they trace large scale structures by exploiting the deep
radio data available for $Ks$-band survey. We find that only a small
fraction of $Ks<20$\,mag EROs have  radio counterparts to the $\rm
60\mu Jy$ limit of the radio data (17 out of 289). This is much lower
than the identification fraction reported by Smail et al. (2002;
21/68) and Roche et al. (2003; 7/31) which is attributed  to the
deeper radio observations in those studies. Indeed, Smail et
al. (2002)  find that the  number of EROs detected at radio
wavelengths rapidly increases at  faint flux densities, doubling below
$\approx \rm 40\mu Jy$. Using stacking analysis we estimate a mean
radio flux density for EROs of $\approx 8\,\mu$Jy, about 1\,dex
lower than the limit of our survey. Much deeper observations are
therefore required to identify the bulk of the radio counterparts for
EROs.

Cross-correlating the radio and ERO positions (after excluding the 17
EROs with 1.4\,GHz emission) does not produce any signal, suggesting
little association between the two populations. However,  for the
sub-sample of radio sources with photometric redshifts $z>0.6$ or no
optical counterparts (assumed to lie at high-$z$) we get a marginally
significant signal at the $\approx2\sigma$ level. Although the small
number statistics (only 43 radio sources fulfilling the above
criteria) hamper a secure interpretation, the evidence above  suggests
that it is the high-$z$ radio population that appears to trace similar
large scale structures with EROs.

A large fraction of the radio population however, to the limit of our
survey, is associated with moderate and low-$z$ starburst systems
(Georgakakis et al. 1999; Afonso et al. 2004) tracing structures at 
$z<1$. Cross-correlation of the full radio sample with the $Ks$-band
density maps gives a statistical significant signal ($\ga 3\sigma$)
only at $z\la0.7$, contrary to the ERO population with a
cross-correlation signal that peaks at $z\ga1$. This explains why
cross-correlation of the full radio and ERO samples does not produces
a statistically significant signal.   

Summarising our conclusions:

\begin{itemize}

\item using angular correlation function analysis we estimate a
statistically  significant signal ($>3\sigma$) for $Ks<20$\,mag
EROs. The derived correlation function amplitude is consistent with
previous studies that used smaller sample sizes. This amplitude
translates to clustering lengths in the range $r_{o}=12-17\,h^{-1} \rm 
\,Mpc$. 

\item cross-correlation of the ERO positions with the $Ks$-band
density maps gives a statistically significant signal only for the
$z\ga1$ maps. This cross-correlation signal is higher, albeit at the
$2-2.5\sigma$ level, from that obtained for non-ERO galaxies with the
same magnitude distribution as our ERO sample. We argue that this is
direct evidence  that EROs are associated with regions of enhanced
density at redshifts  $z>1$. 

\item 17 of the 289 EROs with $K<20$\,mag show radio emission with
flux densities in the range $\rm 65-1000\mu Jy$. Using stacking
analysis we estimate a mean radio flux density of $\approx \rm 8\,
\mu$Jy for EROs.

\item cross-correlation of the radio and the ERO samples (after
excluding the 17 EROs  with radio counterparts) gives a $2\sigma$
signal only for the sub-sample of high-$z$ radio sources
($z\ga0.6$). Although the  statistics are poor this suggests that the
bulk of the radio and the ERO populations trace different
structures. Indeed, cross-correlation of the radio positions with the
$Ks$-band density maps gives a statistically significant signal only
for the $z\la0.7$ maps, contrary to EROs. 

\end{itemize}

\section{Acknowledgments}
 We thank the anonymous referee for useful comments and suggestions. 
 AG acknowledges funding by the European Union and the Greek Ministry
 of Development in the framework of the Programme 'Competitiveness-
 Promotion of Excellence in Technological Development and Research-
 Action 3.3.1', Project 'X-ray Astrophysics with ESA's mission XMM',
 MIS-64564. AMH acknowledges support provided by the National
 Aeronautics and Space Administration through Hubble Fellowship grant
 HST-HF-01140.01-A awarded by the Space  Telescope Science Institute.
 JA gratefully acknowledges the support from the Science and
 Technology Foundation (FCT, Portugal) through the fellowship
 BPD-5535-2001 and the research grant POCTI-FNU-43805-2001. The
 Phoenix Deep Survey radio data are electronically available at {\sf 
  http://www.atnf.csiro.au/people/ahopkins/phoenix/}.

\end{document}